\documentclass[conference]{IEEEtran}
\IEEEoverridecommandlockouts
\usepackage{cite}
\usepackage{graphicx}
\usepackage{subfigure}
\usepackage{amsmath,amssymb,amsfonts}
\usepackage{algorithmic}
\usepackage{textcomp}
\usepackage{xcolor}
\usepackage{bm}
\def\BibTeX{{\rm B\kern-.05em{\sc i\kern-.025em b}\kern-.08em
    T\kern-.1667em\lower.7ex\hbox{E}\kern-.125emX}}
\begin{document}

\title{Preliminary Design of a FADC Readout System for the Alpha/Beta Discrimination in a Large Area Plastic Scintillation Detector}
\author{Jingjun Wen, Daowei Dou, Jinfu Zhu, Zhi Zeng, Tao Xue*\footnote{Corresponding author is Tao Xue, e-mail: xuetaothu@mail.tsingua.edu.cn}, Jianmin Li, Junli Li, Yinong Liu}
\thanks{This work was supported by the National Key Research and Development Program of China (2017YFA0402202)).}
\thanks{The authors are with the Department of Engineering Physics, Tsinghua University, and also with the Key Laboratory of Particle and Radiation Imaging (Tsinghua University), Ministry of Education, Beijing, 100084, China.(Corresponding author is Tao Xue, e-mail: xuetaothu@mail.tsingua.edu.cn).}

\maketitle

\begin{abstract}
This paper describes a FADC Readout system developed for the tap water $\alpha/\beta$ dose monitoring system which is based on EJ444 phoswich scintillation detector and wavelength shifting fiber readout. The Readout system contains dual sampling channels that can supply sampling rate up to 1 GSPS, 14-Bit vertical resolution and adequate effective number of bits (9.7 Bits at 10 MHz), which is optimum for the discrimination of the minimal difference between alpha/beta signals. Moreover, the system is based on a ZYNQ SoC which provides high data throughput speed, low latency and excellent flexibility. As for the discrimination algorithms, a simple least-square classification method is used to discriminate the $\alpha/\beta$ signals and shows high discrimination capability. Besides, the front-end electronics and high voltage supply modules are also briefly introduced in the paper.
\end{abstract}


\section{Introduction}
The safety of tap water is important in the civil security and is very sensitive that have to be protected against the radioactive leak and terrorist attacks. Real-time monitoring of the total $\alpha/\beta$ dose of tap water plays an essential role in ensuring the radioactive safety of tap water plant\cite{ref1}. We have designed a large area alpha/beta detector which combined the EJ444 phoswich scintillators and WLS fibers. The EJ444 scintillator is a combination of EJ440 (ZnS:Ag) and EJ212 (plastic scintillator), where the ZnS:Ag is sensitive to the $\alpha$ particles while EJ212 can detect the $\beta$/$\gamma$-rays. Moreover, due to the differences between the waveforms from ZnS:Ag and EJ212, the PSD methods can be applied to discriminate the $\alpha/\beta$ signals. 

In this paper, the Readout and trigger system based on FADC designed for the real-time alpha/beta monitoring is described. The Readout system provides two sampling channels with 1 GSPS sampling rate and 14-Bit vertical resolution. Especially, the ENOB of this Readout system can achieve $\sim$10 Bits which can show more detailed information of the waveforms from detector and help us to research its performance and properties. Besides, the Readout system transfers the data through the Gigabit Ethernet and adopts the ping-ping readout to reduce the dead time, which makes it suitable for high count rate measurement situation. 

Moreover, the off-line data process algorithms are also studied for the real-time analysis with an industrial computer. A simple least square classification algorithm is used to discriminate the alpha/beta signals, and the real-time analyzed algorithm provides an optimum discrimination efficiency. 

\section{Methods}
\subsection{Structure of the detector and readout electronics}
\begin{figure}[htbp]
\includegraphics[width = 0.5\textwidth]{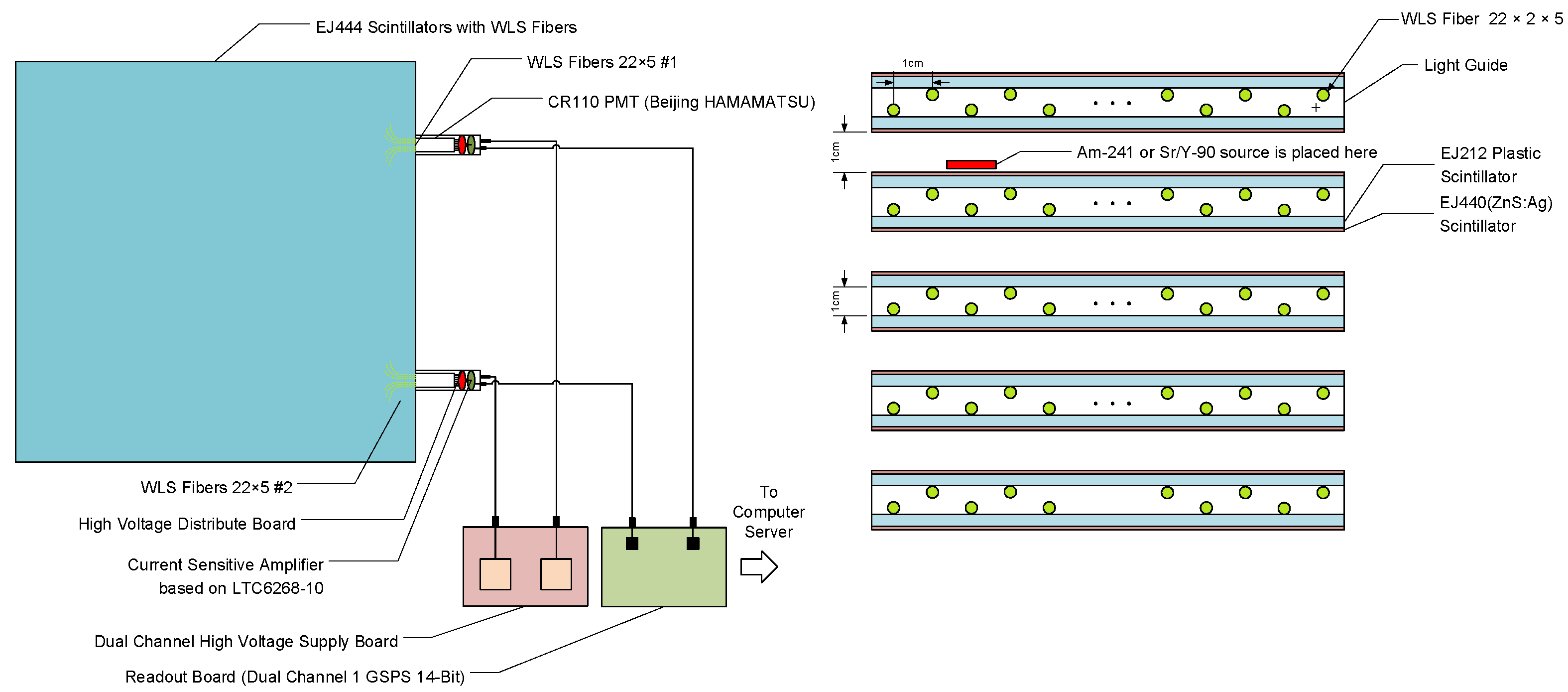}
\caption{The structure of the detector and corresponding electronics modules}
\label{fig:1}
\end{figure}

The structure of the detector and corresponding electronics are shown in fig.\ref{fig:1}. In this detector, the WLS fibers are placed in the grooves on a light guide and the light guide is sandwiched between two pieces EJ444 scintillators. These structures are combined to create a detector module, and the alpha/beta detector in this paper consists of 5 modules.  

\begin{figure}[htbp]
\centering
\subfigure[]{
\label{fig2:a}
\includegraphics[width = 0.4\textwidth]{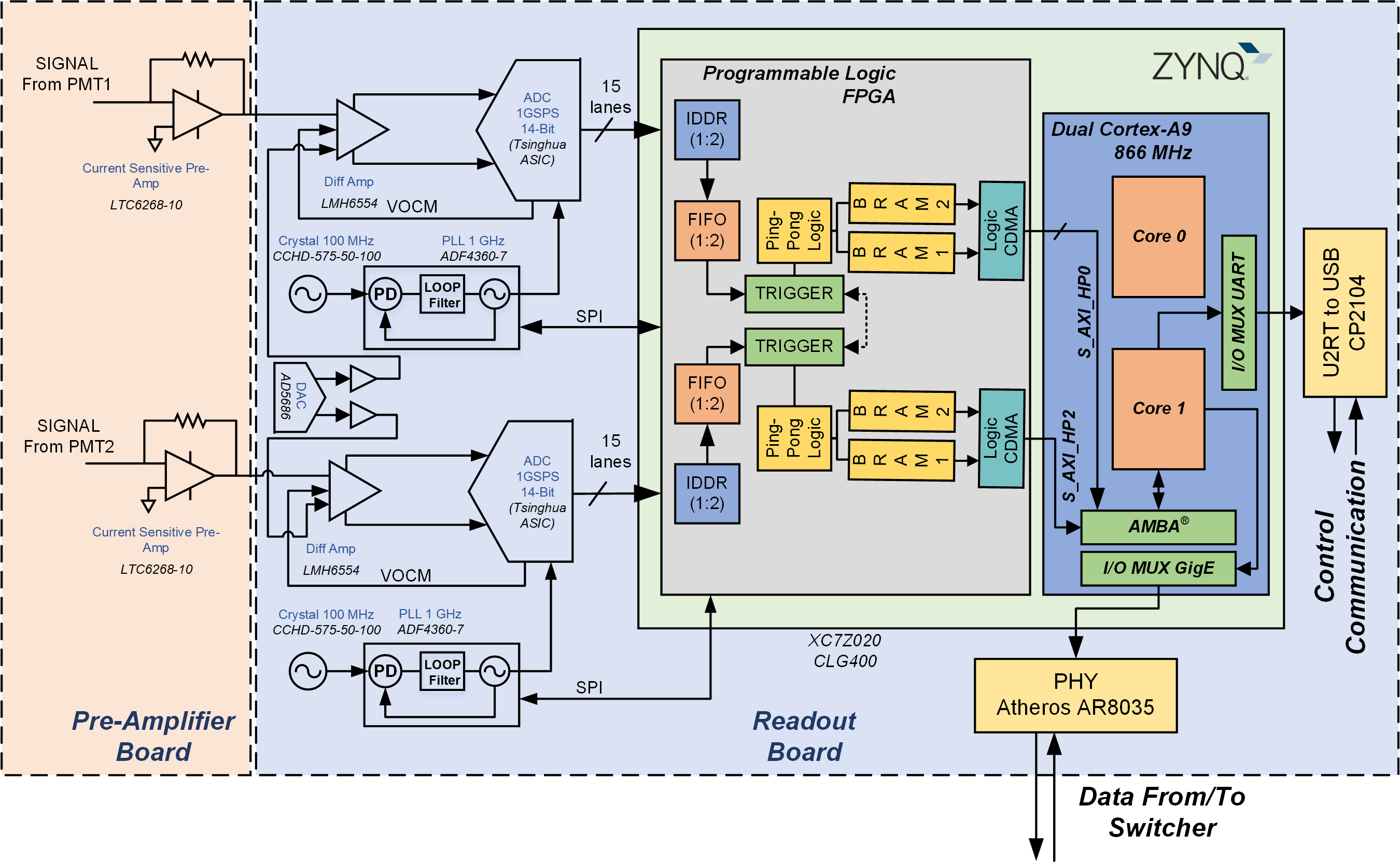}}
\subfigure[]{
\label{fig2:b}
\includegraphics[width = 0.21\textwidth]{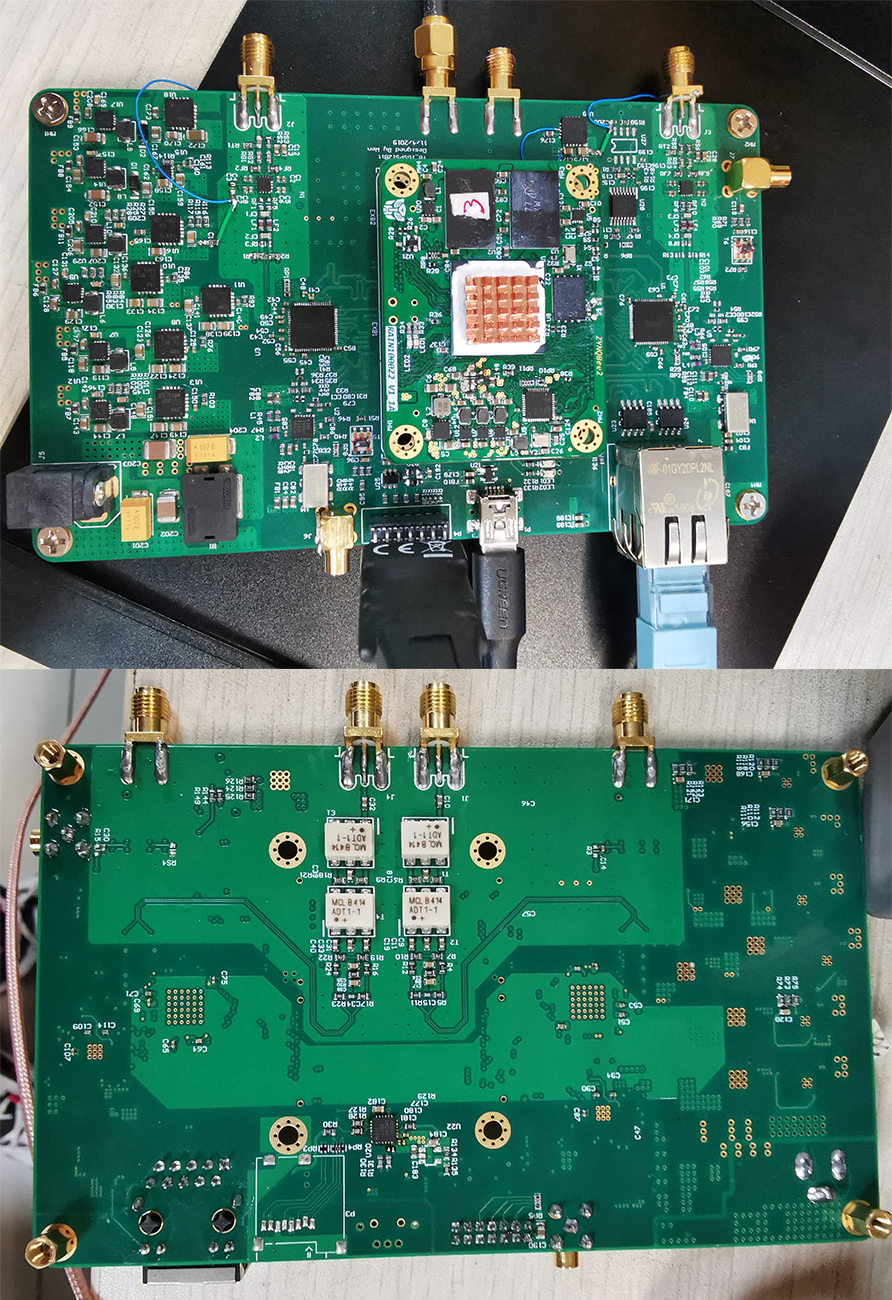}}
\subfigure[]{
\label{fig2:c}
\includegraphics[width = 0.24\textwidth]{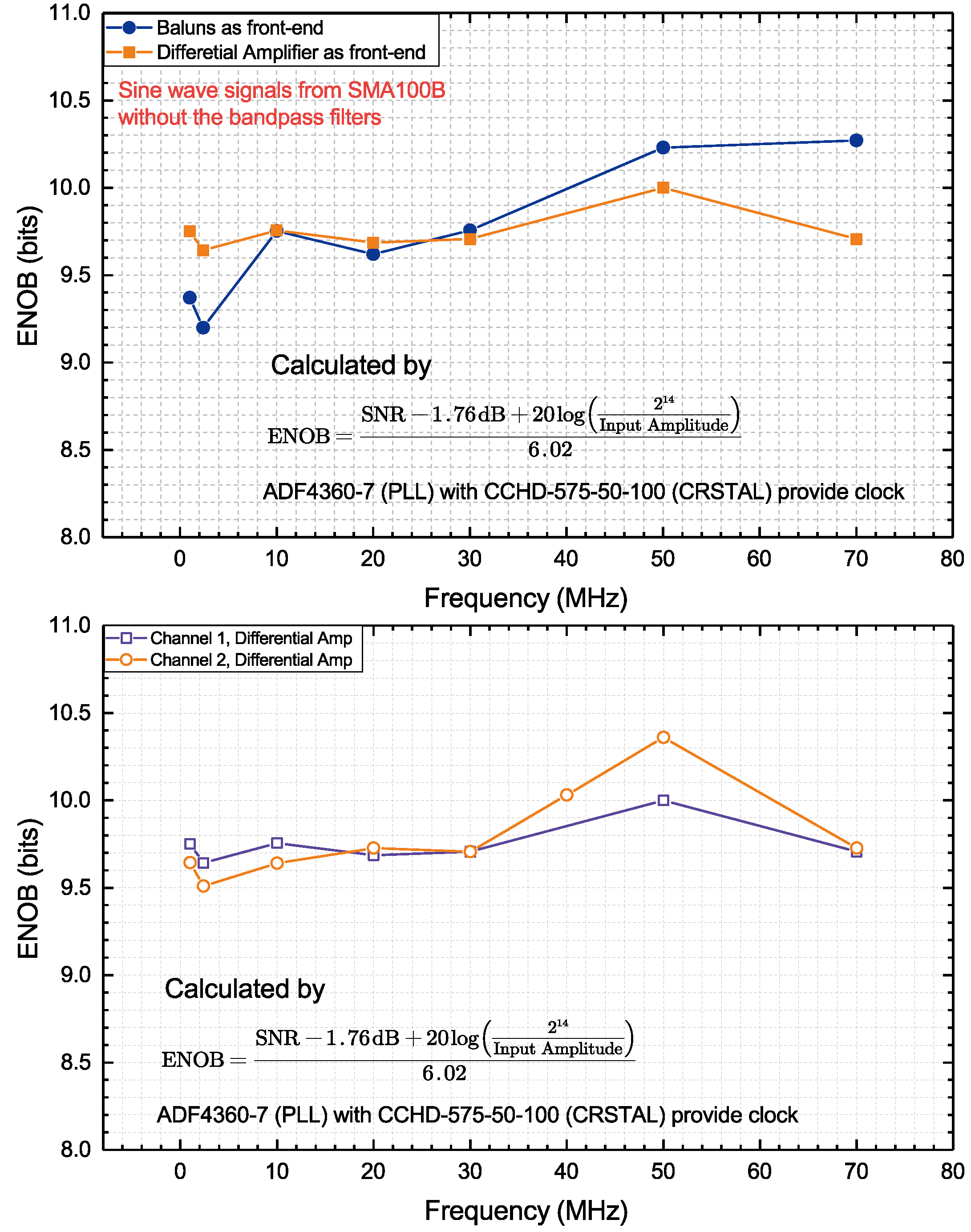}
}
\caption{(a) Detailed structure of the readout board, (b) The photo of readout board, (c) The ENOB, DNL and INL measurement results for the readout board.}
\end{figure}

The readout board is based on two TLG121G ADCs (1 GSPS, 14-Bit) and ZYNQBee 2 readout module\cite{ref2}. It can provide the high-sampling rate (1 GSPS), optimum ENOB (9.7@10 MHz, measured as IEEE 1241-2000 standard) and low dead time. The detailed structure and parameters are shown in Fig.2. As described in figure \ref{fig2:a}, the signals from the anode of the PMT are firstly amplified by the current sensitive pre-amplifier and enter the differential amplifier. Then the differential analog signals are digitized by 1 GSPS, 14-Bit TLG2121G ADCs (Tsinghua ASIC), and the digitized signals are processed by a ZYNQ XC7Z020-CLG400 SoC chip which provides both programmable logic (similar to the Artix-7 FPGA) and two Cortex-A9 processors. The firmware design on ZYNQ SoC includes PLL, DAC configuration via SPI, ADC data buffering, trigger and other logical design. The embedding Linux operation system is running on the Cortex-A9 processor to establish the TCP/IP connection with the server computer, and waveform data are sent through 1000BASE-T gigabit ethernet.

Besides, the high voltage board is based on two CC228P-01Y HV modules (Beijing HAMAMATSU) and can supply voltage up to 1.2 kV. The front-end electronics adopts a current-sensitive preamplifier structure based on LTC6268-10 with a trans-impedance gain of 10 k$\Omega$.
\subsection{Experiment Setup}
The Am-241 and Sr/Y-90 solid radioactive sources are used to provide alpha/beta signals, and they are individually placed in the gaps of two detector modules for alpha/beta measurement experiments. Moreover, the background measurement experiment with a Pb shield is also carried out. In the experiments, the integration time for one hit is 1.2 $\mu$s and more than 500,000 hits were recorded in each experiment.
\subsection{Algorithms}
In this paper, least-square classification (LS) is used as the PSD algorithm\cite{ref3}. Given a training dataset $\mathrm{D}=\{(\bm{x}_1,y_1),\cdots,(\bm{x}_n,y_n)\}$, where $\bm{x}_i = (x_{i1},\cdots,x_{id})$ is the waveform from EJ444 detector and $y_i\in\{-1,+1\}$ represents the waveform is generated by $\alpha$ particle ($y=+1$) or $\beta$ particle ($y=-1$). The target of LS is to find a weight vector $\bm{w}$ and bias $b$, such that $\bm{w}^T\bm{x}_i+b\approx y_i$. 
\section{Results}
\subsection{The $\alpha/\beta$ discrimination results}
\begin{figure}[htbp]
\centering
\includegraphics[width = 0.35\textwidth]{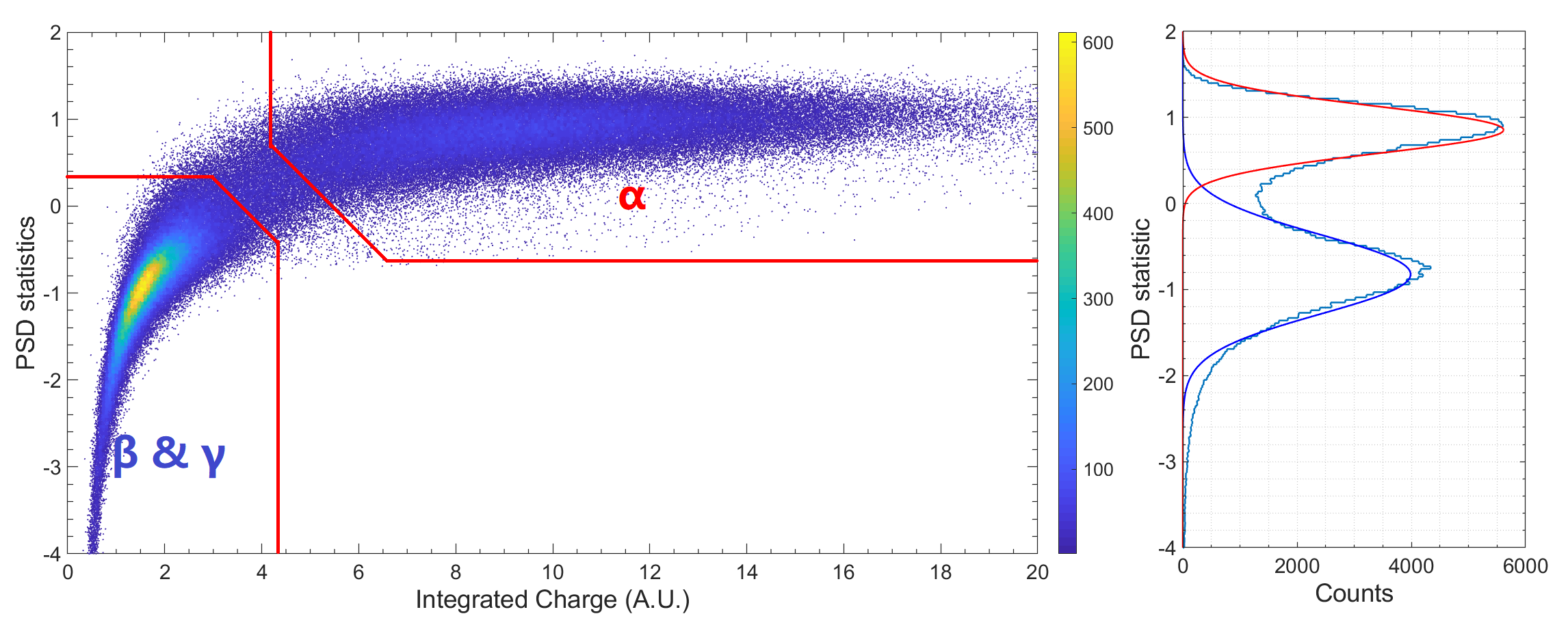}
\caption{The discrimination plot: distribution of waveforms based on their LS PSD statistic versus the integral values}
\label{fig:3}
\end{figure}

\begin{figure}[htbp]
\centering
\includegraphics[width = 0.4\textwidth]{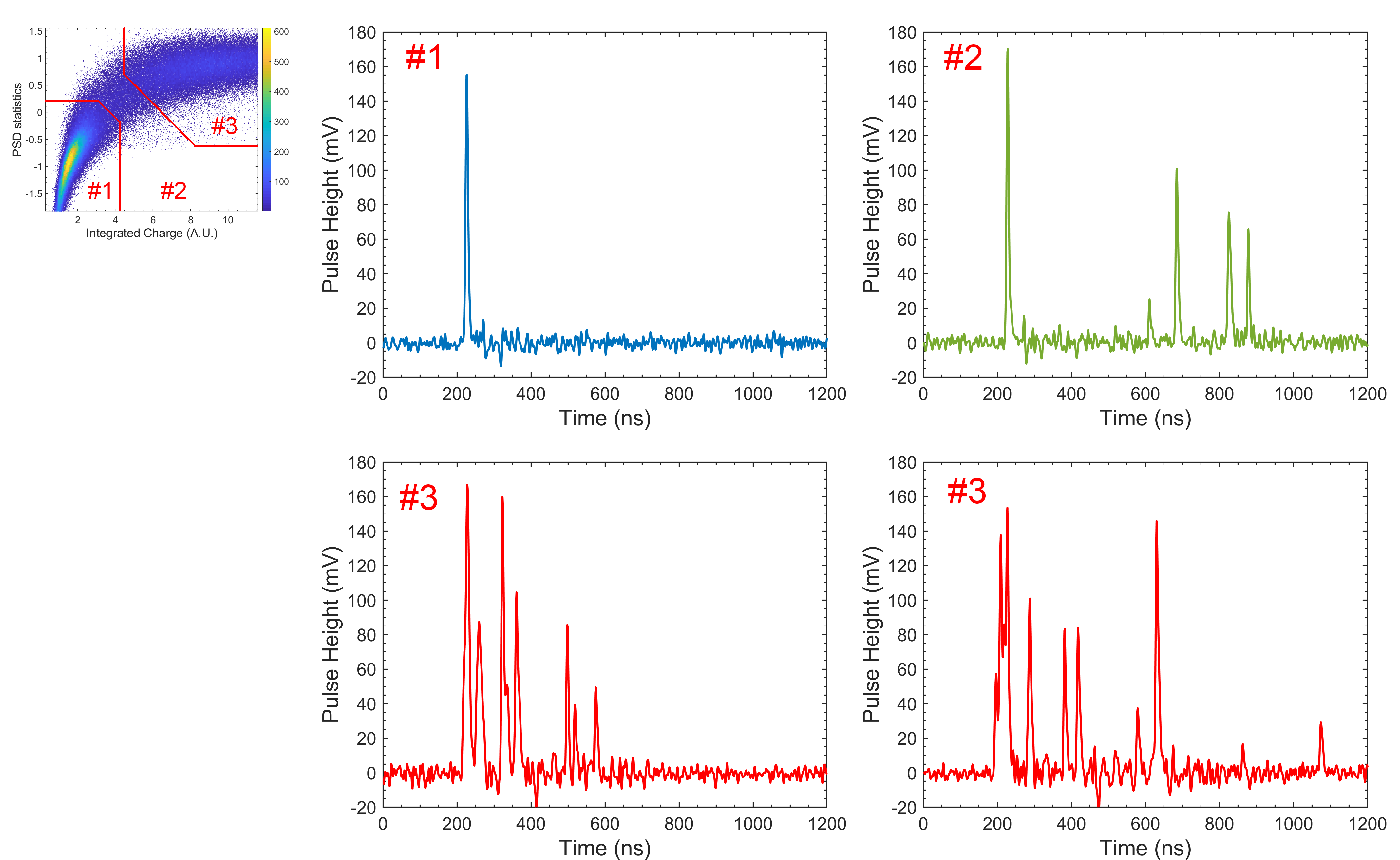}
\caption{The typical waveforms in part \#1, \#2 and \#3.}
\label{fig:4}
\end{figure}

Baselines of all waveforms are subtracted by letting the entire waveform minus the average baseline value, and then the data from the experiment with Am-241 are selected as the dataset which is going to be discriminated by LS algorithm. The discrimination results are shown in figure \ref{fig:3}, the x-axis represents the integral of the waveforms (integral interval is 0$\sim$1200 ns), and the y-axis is the discrimination statistic obtained by LS.
From the figure \ref{fig:3}, it's clearly that there is a intermediate zone between $\alpha$ signals and $\beta/\gamma$ signals. In order to research the pulse shape in this zone, the 2D distribution in figure \ref{fig:3} is divided into 3 parts: \#1, \#2 and \#3. The typical waveforms of the 3 parts are shown in figure \ref{fig:4}.
\subsection{The determination of $\alpha/\beta$ signals' ROI}
The charge integral spectrum from Am-241, Sr/Y-90 and background experiments are shown in figure 5.
\begin{figure}[htbp]
\centering
\includegraphics[width = 0.32\textwidth]{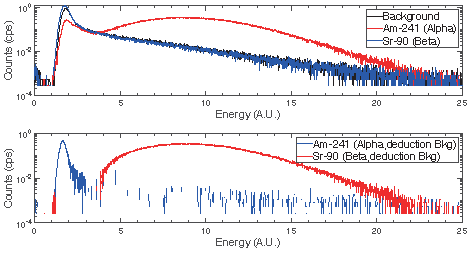}
\caption{Charge integral spectrum of Am-241, Sr/Y-90 and background}
\label{fig:5}
\end{figure}
\section{Summary}
In this paper, a readout system which can provide two 1 GSPS, 14-Bit sampling channels is designed to discriminate the $\alpha/\beta$ signals in a tap water $\alpha/\beta$ dose monitoring detector. The structure and performance are described briefly, and the discrimination results with typical $\alpha/\beta$ waveforms are also shown. 
\section*{Acknowledgment}
We would like to thank those who collaborated on the CDEX, and also thankwords of deep appreciation go to Professor Yulan Li, Qian Yue, Litao Yang and Guang Meng for their invaluable advices, supports and various discussions over the years at the Tsinghua University DEP (Department of Engineering Physics).
	 
We are grateful for the patient help of Yu Xue, Wenping Xue, and Jianfeng Zhang. They are seasoned, full-stack hardware technologists with rich experience of solder and rework in the electronics workshop at DEP.

\end{document}